%% file: main.tex
\documentclass[preprint,12pt]{elsarticle}
\usepackage{amsmath,amssymb,mathtools,amsthm}
\usepackage{booktabs,tabularx,array}
\usepackage{enumitem}
\usepackage{microtype}
\usepackage{graphicx}
\usepackage{tikz}
\usetikzlibrary{arrows.meta,positioning,shapes.geometric}
\usepackage{xcolor}
\usepackage{listings}
\usepackage{hyperref}
\usepackage[nameinlink,noabbrev]{cleveref}
\usepackage{url}
\input{lstlean.tex}
\newcommand{\leanin}[1]{\texttt{\color{identcolor}#1}}

\definecolor{keywordcolor}  {HTML}{0000FF}  
\definecolor{tacticcolor}   {HTML}{0000FF}  
\definecolor{sortcolor}     {HTML}{267F99}  
\definecolor{commentcolor}  {HTML}{008000}  
\definecolor{symbolcolor}   {HTML}{A31515}  
\definecolor{attributecolor}{HTML}{795E26}  
\definecolor{springcolor}   {HTML}{001080}  

\hypersetup{
  hidelinks,
  pdftitle={A Proof of the Dittert Conjecture in Dimension 4 via an Exact Constrained Sum-of-Squares Certificate},
  pdfauthor={Jinhui Li, Beibei Xiong and Zhengfeng Yang}
}

\setlist[itemize]{leftmargin=2em,itemsep=0.2em,topsep=0.35em}
\setlist[enumerate]{leftmargin=2.2em,itemsep=0.2em,topsep=0.35em}

\newtheorem{theorem}{Theorem}[section]

\newtheorem{conjecture}[theorem]{Conjecture}

\theoremstyle{definition}

\newcommand{\per}{\operatorname{per}}
\newcommand{\R}{\mathbb{R}}

\newcommand{\Frob}[1]{\left\lVert #1\right\rVert_{\mathrm F}}

\journal{Journal of Symbolic Computation}

\begin{document}

\begin{frontmatter}

\title{A Proof of the Dittert Conjecture in Dimension 4 via an Exact Constrained Sum-of-Squares Certificate}

\author[ecnu]{Jinhui Li}
\ead{jinhuili@stu.ecnu.edu.cn}
\author[ecnu]{Beibei Xiong}
\ead{bbxiong@stu.ecnu.edu.cn}
\author[ecnu]{Zhengfeng Yang}
\ead{zfyang@sei.ecnu.edu.cn}

\address[ecnu]{Shanghai Key Laboratory of Trustworthy Computing, 
East China Normal University, Shanghai 200062, China}

\begin{abstract}
The Dittert conjecture states that the Dittert functional on
nonnegative \(n\times n\) matrices whose entries sum to \(n\) is
uniquely maximized by the uniform matrix.  We prove the conjecture in
dimension \(4\).  More precisely, let \(K_4\) be the simplex of
nonnegative \(4\times4\) real matrices whose entries sum to \(4\), let
\(U_4\) be the uniform matrix, and let \(\phi\) denote the Dittert
functional.  We establish
\(\frac{61}{32}-\phi(A)\geq
\frac1{52}\lVert A-U_4\rVert_F^2\) for every \(A\in K_4\).
Consequently, \(U_4\) is the unique maximizer of \(\phi\) on \(K_4\).

 The proof reduces the problem to an exact certification of the
nonnegativity of a structured quartic polynomial in sixteen variables on a
simplex. We develop a symbolic--numeric procedure for constructing an
exact rational constrained sum-of-squares certificate. The procedure
combines adaptive template selection with sequential rational recovery to
handle singular Gram matrices and coupled SOS blocks arising from the
constraint structure. The final certificate consists of a main SOS with
\(152\) positively weighted rational squares and \(136\) smaller SOS blocks,
each containing \(16\) such squares. Exact \(LDL^{\mathsf T}\) decompositions
and coefficient comparison over \(\mathbb Q\) certify the polynomial
identity. The resulting exact certificate is formally verified using the
Lean proof assistant.
\end{abstract}


\end{frontmatter}

\section{Introduction}\label{sec:introduction}

The permanent is a classical matrix invariant in linear algebra and
combinatorics.  Its extremal behavior on nonnegative matrices includes the
van der Waerden theorem for doubly stochastic matrices.  The Dittert
conjecture asks for an analogous extremal characterization on the larger
simplex obtained by fixing only the sum of all matrix entries: it asserts that
the uniform matrix uniquely maximizes a functional formed from the products
of the row sums, the products of the column sums, and the permanent.  

The conjecture was recorded as Conjecture~28 in Minc's survey
\cite{Minc1987}; Hajek proposed a related generalized permanent inequality
motivated by a multiaccess communication problem
\cite{Hajek1987,CheonWanless2012}.  
The cases \(n=2\) and \(n=3\) were proved by Sinkhorn and Hwang,
respectively
\cite{Sinkhorn1984,Hwang1986,Hwang1987}.  Cheon and Wanless obtained
further restrictions on possible maximizers
\cite{CheonWanless2012}.  More recently, Udayan and Somasundaram proved
that every maximizing
matrix in the four-dimensional Dittert problem is fully indecomposable
\cite{UdayanSomasundaram2024}. 
Pang presented a proof of the conjecture
for \(n\geq17\) in a recent arXiv preprint \cite{Pang2026}.  In this paper, we prove the complete
dimension-\(4\) statement.

Let \(K_4\) denote the simplex of nonnegative
\(4\times4\) matrices whose entries sum to \(4\). 
For
\(A\in K_4\), let \(r_i(A)\) and \(c_j(A)\) denote its row and column
sums, respectively, and define
\[
  \phi(A)
  :=
  \prod_{i=1}^{4}r_i(A)
  +
  \prod_{j=1}^{4}c_j(A)
  -
  \per(A),
\]
where \(\per(A)\) denotes the permanent of $A$. 
Our first main result is the following.

\begin{theorem}[Dittert conjecture in dimension $4$]\label{thm:main}
For every $A\in K_4$,
\[
  \phi(A)\leq\frac{61}{32}.
\]
Moreover,
\[
  \phi(A)=\frac{61}{32}
  \quad\Longleftrightarrow\quad
  A=U_4.
\]
\end{theorem}

We prove this theorem through the following stronger estimate.

\begin{theorem}[Quantitative stability in dimension $4$]\label{thm:stability}
For every $A\in K_4$,
\begin{equation}\label{eq:phi-stability}
  \frac{61}{32}-\phi(A)
  \geq
  \frac1{52}\Frob{A-U_4}^{2}.
\end{equation}
\end{theorem}
The proof of Theorem~\ref{thm:stability}  reduces to establishing
the nonnegativity of a structured
quartic polynomial in sixteen variables on a simplex,
which has affine dimension \(15\). 
Semidefinite programming provides an
effective approach for discovering candidate SOS representations. However,
the resulting floating-point certificates are only numerical
approximations and do not constitute exact proofs.

Building on symbolic--numeric SOS recovery methods
\cite{PeyrlParrilo2008,KaltofenLiYangZhi2012}, we consider a more
challenging exact certification problem arising from a constrained SOS
representation. Unlike the settings addressed by existing recovery
techniques, our problem involves singular Gram matrices induced by
the equality point \(U_4\) and multiple coupled Gram blocks linked by a
linear equality constraint. These structural features prevent a direct application of standard
rational recovery procedures.

To overcome these difficulties, we develop an adaptive search strategy
combined with a sequential recovery procedure for constructing the exact
constrained SOS certificate. The
procedure searches a hierarchy of certificate templates and performs
sequential rational recovery, i.e.,  smaller Gram blocks are recovered and fixed
first, followed by the reconstruction of the reduced main Gram matrix.
The resulting certificate contains a main SOS with \(152\) positively
weighted rational squares and \(136\) smaller SOS blocks, each consisting of
\(16\) such squares. Exact \(LDL^{\mathsf T}\) decompositions certify
positivity, and the polynomial identity is verified over \(\mathbb Q\) and
formally checked in Lean. 
The numerical SDP computation is used only to discover candidate
certificate structures, while the final certificate is established by exact
symbolic verification.

The rest of the paper is organized as follows.
\Cref{sec:preliminaries} describes the polynomial formulation of the Dittert
conjecture in dimension $4$.  
We introduce the
quantitative stability problem and the SOS relaxations 
in \Cref{sec:stability-sos}.
We present the template search and the
sequential exact recovery of the certificate in \Cref{sec:agent-certificate}.  Finally, \Cref{sec:lean-formalization} presents the Lean formalization
of the exact certificate.

\section{Preliminaries and Polynomial Formulation}\label{sec:preliminaries}

\subsection{The permanent and the Dittert conjecture}

For \(A=[a_{ij}]\in\R^{n\times n}\), define its permanent by
\begin{equation}\label{eq:permanent}
  \per(A)
  :=
  \sum_{\sigma\in\mathfrak S_n}
  \prod_{i=1}^{n}a_{i,\sigma(i)},
\end{equation}
where \(\mathfrak S_n\) denotes the symmetric group on
\(\{1,\ldots,n\}\).  We write
\[
  r_i(A):=\sum_{j=1}^{n}a_{ij},
  \qquad
  c_j(A):=\sum_{i=1}^{n}a_{ij}
\]
for the \(i\)-th row sum and \(j\)-th column sum, respectively. Let
\[
  \Omega_n
  :=
  \left\{
    A\in\R_{\ge0}^{n\times n}:
    r_i(A)=1\ \text{for all }i,
    \quad
    c_j(A)=1\ \text{for all }j
  \right\}
\]
denote the set of $n\times n$ doubly stochastic matrices.  Let $U_n$
denote the uniform matrix whose entries are all equal to $1/n$.
The van der Waerden theorem
\cite{Egorychev1981,Falikman1981} states that
\begin{equation}\label{eq:vdw}
  \per(A)\geq\frac{n!}{n^n},
  \qquad A\in\Omega_n,
\end{equation}
with equality if and only if \(A=U_n\).

The Dittert problem is posed on the nonnegative matrix simplex
\begin{equation}\label{eq:Kn}
  K_n
  :=
  \left\{
    A\in\R_{\ge0}^{n\times n}:
    \sum_{i=1}^{n}\sum_{j=1}^{n}a_{ij}=n
  \right\}.
\end{equation}
Clearly, $\Omega_n\subseteq K_n$. For $A\in K_n$, define the Dittert
functional by
\begin{equation}\label{eq:phi}
  \phi(A)
  :=
  \prod_{i=1}^{n}r_i(A)
  +
  \prod_{j=1}^{n}c_j(A)
  -
  \per(A).
\end{equation}
At the uniform matrix,
\[
  \per(U_n)=\frac{n!}{n^n},
  \qquad
  \phi(U_n)=2-\frac{n!}{n^n}.
\]

\begin{conjecture}[Dittert]\label{conj:dittert}
For every integer \(n\geq 2\) and every \(A\in K_n\),
\begin{equation}\label{eq:dittert-conjecture}
  \phi(A)
  \leq
  \phi(U_n)
  =
  2-\frac{n!}{n^n},
\end{equation}
with equality if and only if \(A=U_n\).
\end{conjecture}
We denote the assertion of Conjecture~\ref{conj:dittert} in dimension \(n\) by
\(DH_n\).  On \(\Omega_n\), one has
\[
  r_i(A)=c_j(A)=1
\]
for all \(i,j\), and hence
\[
  \phi(A)=2-\per(A).
\]
Thus, the restriction of the Dittert conjecture to \(\Omega_n\) is precisely
the van der Waerden lower bound for the permanent.

\subsection{The dimension-four polynomial formulation}
We write \(K_4\) for the case \(n=4\) of the matrix simplex
defined in \eqref{eq:Kn}.
For $x=(x_1,\ldots,x_{16})\in\R^{16}$, set
\begin{equation}\label{eq:A-of-x}
  A(x)
  :=
  \begin{pmatrix}
    x_1  & x_2  & x_3  & x_4  \\
    x_5  & x_6  & x_7  & x_8  \\
    x_9  & x_{10} & x_{11} & x_{12} \\
    x_{13} & x_{14} & x_{15} & x_{16}
  \end{pmatrix}.
\end{equation}
Then $A(x)\in K_4$ if and only if
\begin{equation}\label{eq:K4-coordinates}
  x_i\ge0\quad(1\le i\le16),
  \qquad
  \sum_{i=1}^{16}x_i=4.
\end{equation}

Since
\[
  \per(U_4)=\frac{4!}{4^4}=\frac{3}{32},
  \qquad
  \phi(U_4)=2-\frac{3}{32}=\frac{61}{32},
\]
we define the quartic integer-coefficient gap polynomial
\begin{equation}\label{eq:Q4}
\begin{aligned}
  Q_4(x)
  &:=
  32\bigl(\phi(U_4)-\phi(A(x))\bigr)\\
  &=
  61
  -32\prod_{i=1}^{4}r_i\bigl(A(x)\bigr)
  -32\prod_{j=1}^{4}c_j\bigl(A(x)\bigr)
  +32\per\bigl(A(x)\bigr).
\end{aligned}
\end{equation}
Whenever \(A=A(x)\), we also write \(Q_4(A)\) for \(Q_4(x)\).
Consequently, the conjecture \(DH_4\) is equivalent to
\begin{equation}\label{eq:DH4-polynomial}
  Q_4(A)\geq0
  \qquad\text{for all }A\in K_4,
\end{equation}
with equality if and only if \(A=U_4\).
The next section formulates a quantitative stability strengthening of
\eqref{eq:DH4-polynomial}.

\section{Quantitative Stability and SOS Relaxations}
\label{sec:stability-sos}

We formulate the stability problem as an exact constrained sum-of-squares certification problem.
To obtain such certificates, we introduce degree-bounded pairwise preordering relaxations.

\subsection{A quantitative stability estimate}

For \(A=A(x)\) as in \eqref{eq:A-of-x}, define
\begin{equation}\label{eq:Delta4}
  \Delta_4(A)
  :=
  \Frob{A-U_4}^{2}
  =
  \sum_{k=1}^{16}
  \left(x_k-\frac14\right)^2,
\end{equation}
where \(\Frob{\cdot}\) denotes the Frobenius norm.  We also write
\(\Delta_4(x):=\Delta_4(A(x))\).

We seek an explicit constant \(\kappa>0\) such that
\begin{equation}\label{eq:positive-stability}
  \phi(U_4)-\phi(A)
  \geq
  \kappa\Delta_4(A)
  \qquad
  \text{for all }A\in K_4.
\end{equation}
The existence of the positive constant $\kappa$  proves \(DH_4\), including uniqueness of the equality
case, since \(\Delta_4(A)=0\) if and only if \(A=U_4\).  
We call any constant \(\kappa>0\) satisfying
\eqref{eq:positive-stability} a stability constant for \(DH_4\).

For \(\lambda\in\R\), define
\begin{equation}\label{eq:parametric-polynomial}
  P_{4,\lambda}(x)
  :=
  Q_4(x)-\lambda\Delta_4(x).
\end{equation}
By \eqref{eq:Q4}, the condition
\begin{equation}\label{eq:parametric-nonnegativity}
  P_{4,\lambda}(x)\geq0
  \qquad
  \text{for }A(x)\in K_4
\end{equation}
is equivalent to
\[
  \phi(U_4)-\phi(A)
  \geq
  \frac{\lambda}{32}\Delta_4(A)
  \qquad
  \text{for all }A\in K_4.
\]
It therefore suffices to certify
\eqref{eq:parametric-nonnegativity} for some \(\lambda>0\).

\subsection{A truncated pairwise preordering}

Denote by $\Sigma\R[x]^2$  the set of all sum-of-squares polynomials,
and set
\begin{equation}\label{eq:simplex-equality}
  h(x):=4-\sum_{i=1}^{16}x_i.
\end{equation}
Under the coordinate identification $A=A(x)$,
\[
  A(x)\in K_4
  \quad\Longleftrightarrow\quad
  x_i\geq0\ (1\leq i\leq16),
  \qquad h(x)=0.
\]
For an integer  $r\geq0$, write $\R[x]_{\leq r}$ for the polynomials of degree at most
$r$, and set
\[
  \Sigma\R[x]^2_{\leq r}
  :=
  \Sigma\R[x]^2\cap\R[x]_{\leq r}.
\]
For an integer \(d\geq2\), define the degree-\(2d\) truncated
pairwise preordering by
\begin{equation}\label{eq:preordering-cone}
  \mathcal T_d
  :=
  \left\{
    \sigma_0
    +\sum_i x_i\sigma_i
    +\sum_{i<j}x_ix_j\sigma_{ij}
    +hq
  \right\},
\end{equation}
where 
  $q\in\R[x]_{\leq 2d-1}$, 
$\sigma_0\in\Sigma\R[x]^2_{\leq2d}$, 
$\sigma_i,\sigma_{ij}\in\Sigma\R[x]^2_{\leq2d-2}$, $1\leq i\leq 16$, $1\leq i <j \leq 16$. 
The set \(\mathcal T_d\) is a convex cone, and every polynomial in
\(\mathcal T_d\) is nonnegative when \(A(x)\in K_4\). 

Define the optimal value of the degree-$2d$ SOS relaxation by
\begin{equation}\label{eq:lambda-d}
  \lambda_d
  :=
  \sup\left\{
    \lambda\in\R:
    P_{4,\lambda}\in\mathcal T_d
  \right\}.
\end{equation}
For our purposes, it suffices to find one positive rational feasible
value of \(\lambda\). For fixed degree bounds, each SOS polynomial has a positive semidefinite
Gram-matrix representation~\cite{Parrilo2000,Lasserre2001},
\begin{equation}\label{eq:gram-representation}
  \sigma(x)=z(x)^{\mathsf T}Gz(x),
  \qquad G\succeq0,
\end{equation}
where $z(x)$ is a prescribed monomial vector. 
Substituting these Gram representations into the expression
for \(\mathcal T_d\), 
\eqref{eq:lambda-d}
can be written as the following semidefinite program
\begin{equation}\label{eq:sdp-relaxation}
  \left\{
  \begin{aligned}
    \text{maximize}\quad
      & \lambda,\\
    \text{subject to}\quad
      & P_{4,\lambda}\in\mathcal T_d.
  \end{aligned}
  \right.
\end{equation}

By setting \(d=2\), the relaxation
\eqref{eq:sdp-relaxation} provides a numerical approach for estimating
the largest feasible stability parameter. The resulting semidefinite
optimization problem yields numerical Gram representations of the SOS
blocks, which serve as candidates for the subsequent exact certificate
construction.

However, the numerical representations obtained from semidefinite
optimization do not constitute an exact proof. To establish the
stability estimate, one needs to recover rational Gram matrices satisfying
the constrained SOS representation exactly. This recovery problem is
particularly challenging because the equality point
\(U_4\) induces singular Gram representations and the constrained certificate
contains multiple coupled Gram blocks.

The next section addresses this exact certificate construction problem by
developing a sequential recovery procedure for singular constrained
sum-of-squares certificates.

\section{Sequential Recovery of Exact Singular Constrained SOS Certificates}
\label{sec:agent-certificate}

Symbolic--numeric methods can recover exact rational SOS decompositions from
numerical Gram representations; see
\cite{PeyrlParrilo2008,KaltofenLiYangZhi2012}.  
However, the certificate
arising from the Dittert conjecture presents additional challenges that are
not directly addressed by standard SOS recovery procedures. First, the equality point \(U_4\) forces every Gram matrix in the original
certificate representation to be singular. Consequently, rational recovery
cannot be applied directly to the original positive semidefinite Gram
matrices. Second, the constrained SOS certificate contains multiple coupled
Gram blocks together with a polynomial multiplier of the simplex equality
\(h=0\). These blocks cannot be recovered independently, since modifying one
block changes the coefficient constraints for the remaining components.

To address these difficulties, we develop a sequential recovery strategy for
constructing the exact constrained SOS certificate. The procedure first
identifies a numerically feasible certificate template and then performs
rational recovery in a prescribed order. The smaller Gram blocks are
recovered and fixed first, after which the reduced main Gram matrix is
reconstructed and corrected so that it remains positive definite. Finally,
the remaining residual is shown to be an exact multiple of the simplex
equality.

The resulting procedure transforms the recovery problem from singular Gram
representations to exact positive definite matrix recovery in reduced
coordinates, and produces a fully rational certificate that can be verified
independently over \(\mathbb{Q}\).

\subsection{Adaptive Search for a Feasible Constrained SOS Template}
\label{subsec:template-search}

The construction of an exact certificate begins with identifying a
constrained SOS template that admits a positive stability parameter. We
consider a prescribed hierarchy of degree-four templates and solve the
corresponding SDP relaxations. The first template yielding a positive
optimal value is selected for the subsequent exact recovery.

The first template in the hierarchy uses only the individual nonnegative
generators \(x_i\):

\begin{equation}\label{eq:first-order-ansatz}
  P_{4,\lambda}
  =
  \sigma_0+\sum_i x_i\sigma_i+hq,
\end{equation}
where
\[
  \sigma_0\in\Sigma\R[x]^2_{\leq4},
  \qquad
  \sigma_i\in\Sigma\R[x]^2_{\leq2},
  \qquad
  q\in\R[x]_{\leq3}.
\]
At the prescribed degree, 
solving the corresponding semidefinite
optimization problem using MOSEK through its Python API
\cite{MOSEKPython} does not yield a positive optimal value for this
template. We therefore extend the template by including
pairwise nonnegative generators \(x_ix_j\):
\begin{equation}\label{eq:successful-template}
  P_{4,\lambda}
  =
  \sigma_0
  +\sum_i x_i\sigma_i
  +\sum_{i<j}x_ix_j\sigma_{ij}
  +hq.
\end{equation}
This is the degree-four certificate template associated with membership in
the truncated pairwise preordering \(\mathcal T_2\) defined in
\eqref{eq:preordering-cone}.  

In this situation, solving the corresponding semidefinite optimization problem gives the numerical optimum
\[
\lambda_2^{\mathrm{num}}\approx0.6153852.
\]
This numerical value motivates the rational choice
\[
\bar\lambda=\frac8{13},
\]
which yields the certified stability constant
\[
\bar\kappa=\frac{\bar\lambda}{32}=\frac1{52}.
\]
The resulting template contains
\(16+\binom{16}{2}=136\) small SOS blocks and one main SOS block. Each
small quadratic SOS uses
\[
  z_{\mathrm{sm}}(x)
  =
  (1,x_1,\ldots,x_{16})^{\mathsf T},
\]
whereas the quartic main SOS uses the vector of all monomials of degree at
most two.  The corresponding numerical Gram matrices have sizes
\(17\times17\) and \(153\times153\), respectively.  These floating-point
matrices determine the candidate certificate structure but do not constitute
an exact proof.

\subsection{Sequential exact recovery of singular constrained SOS certificates}
\label{subsec:exact-recovery}
After obtaining the numerical constrained SOS representation of
\(P_4=P_{4,\bar\lambda}\), the remaining challenge is
to recover an exact rational certificate from the numerical Gram matrices.
The main difficulty
is that the Gram matrices in the original SOS representations are singular
due to the equality point \(U_4\). Let
\[
  c:=\left(\frac14,\ldots,\frac14\right)^{\mathsf T}
\]
be the coordinate vector of \(U_4\). 
Since \(P_4(c)=h(c)=0\) and 
all generators
\(x_i,x_ix_j\) appearing in
\eqref{eq:successful-template} are strictly positive at \(c\),
every SOS block vanishes at
\(c\).  
Hence, for any Gram representation
\[
  \sigma_b=z_b^{\mathsf T}G_bz_b,
  \qquad
  G_b\succeq0,
\] 
one has \(G_bz_b(c)=0\). Since \(z_b(c)\neq0\), every original Gram matrix \(G_b\) is singular.

Ordering each monomial vector with the constant monomial first, write
\[
  z_b(c)
  =
  \begin{pmatrix}
    1\\
    \widetilde z_b(c)
  \end{pmatrix},
\]
and let \(m_b\) be the length of \(z_b\).  We set
\begin{equation}\label{eq:centering-matrix}
  P_b
  :=
  \begin{pmatrix}
    -\widetilde z_b(c)^{\mathsf T}\\
    I_{m_b-1}
  \end{pmatrix},
  \qquad
  L_b
  :=
  (P_b^{\mathsf T}P_b)^{-1}P_b^{\mathsf T}.
\end{equation}
Then \(P_b^{\mathsf T}z_b(c)=0\), and the numerical matrix used for
rational reconstruction is
\begin{equation}\label{eq:reduced-numerical-matrix}
  S_b^{\mathrm{num}}
  :=
  L_bG_b^{\mathrm{num}}L_b^{\mathsf T}.
\end{equation}
All matrices \(S_b^{\mathrm{num}}\) including
\(S_0^{\mathrm{num}}\), are numerically positive definite.  
Thus, the recovery problem is transformed from singular Gram matrices
\(G_b^{\mathrm{num}}\) to positive definite matrices in reduced coordinates, where standard rational recovery techniques can be applied.

The \(136\) small blocks are recovered first, followed by the main block.  For a candidate denominator \(D\), the entries of
\(S_k^{\mathrm{num}}\) are rounded to multiples of \(1/D\), and the
resulting matrix is symmetrized exactly.  Denote this symmetric rational
matrix by \(\widehat S_k^{(D)}\).  The resulting rational matrix is then certified positive definite by an
exact \(LDL^{\mathsf T}\) decomposition.
The candidate denominator is
accepted only if all \(136\) matrices are positive definite.
In this case, the denominator search returns the common
recovery denominator \(D_{\mathrm{sm}}=1331\).
We therefore set
\[
  S_k^{\mathbb Q}
  :=
  \widehat S_k^{(D_{\mathrm{sm}})}
\]
for every small block \(k\).  The corresponding exact Gram matrix and SOS polynomial are
\begin{equation}\label{eq:exact-gram-reconstruction}
  G_k^{\mathbb Q}
  :=
  P_kS_k^{\mathbb Q}P_k^{\mathsf T},
  \qquad
  \sigma_k(x)
  :=
  z_k(x)^{\mathsf T}G_k^{\mathbb Q}z_k(x).
\end{equation}
Since \(S_k^{\mathbb Q}\succ0\) and \(P_k\) has full column rank,
\[
  G_k^{\mathbb Q}\succeq0,
  \qquad
  \ker G_k^{\mathbb Q}
  =
  \operatorname{span}\{z_k(c)\}.
\]
After all \(136\) small blocks have been accepted, they are fixed and
are not modified in the subsequent recovery.
Let
\begin{equation}\label{eq:small-block-contribution}
  C_{\mathrm{sm}}
  :=
  \sum_i x_i\sigma_i
  +
  \sum_{i<j}x_ix_j\sigma_{ij}, 
\end{equation}
 and set
\[
  R_0:=P_4-C_{\mathrm{sm}}.
\]
It remains to find \(\sigma_0\) and \(q\) such that
\[
  R_0=\sigma_0+hq.
\]

For each candidate denominator \(D\), the symbolic computation rationalizes
\(S_0^{\mathrm{num}}\) and applies the exact coefficient corrections,
producing a corrected rational candidate
\(\widehat S_0^{(D)}\).  Set
\[
  \sigma_0^{(D)}
  :=
  \bigl(P_0^{\mathsf T}z_0\bigr)^{\mathsf T}
  \widehat S_0^{(D)}
  \bigl(P_0^{\mathsf T}z_0\bigr),
  \qquad
  E_D:=R_0-\sigma_0^{(D)}.
\]
The corrected candidate is accepted only if
\begin{equation}\label{eq:main-block-requirements}
  \widehat S_0^{(D)}\succ0,
  \qquad
  E_D\left(
    x_1,\ldots,x_{15},
    4-\sum_{i=1}^{15}x_i
  \right)=0.
\end{equation}
The first condition is checked by an exact
\(LDL^{\mathsf T}\) decomposition, and the second coefficientwise in
\(\mathbb Q[x_1,\ldots,x_{15}]\). 

If either condition fails, the recovery parameters are adjusted and the
numerical main block is recomputed if necessary, while retaining the
recovered small blocks. The adaptive denominator search returns the successful denominator
 \(D_0=32767\).  We set
\[
  S_0^{\mathbb Q}
  :=
  \widehat S_0^{(D_0)},
  \qquad
  G_0^{\mathbb Q}
  :=
  P_0S_0^{\mathbb Q}P_0^{\mathsf T},
\]
and
\[
  \sigma_0(x)
  :=
  z_0(x)^{\mathsf T}G_0^{\mathbb Q}z_0(x)
  =
  \bigl(P_0^{\mathsf T}z_0(x)\bigr)^{\mathsf T}
  S_0^{\mathbb Q}
  \bigl(P_0^{\mathsf T}z_0(x)\bigr).
\]
Since \(S_0^{\mathbb Q}\succ0\), the reconstructed matrix
\(G_0^{\mathbb Q}\) is positive semidefinite with corank one.
\(q\)  is obtained by exact polynomial division of
\(R_0-\sigma_0\) by \(h\) and then the remainder is verified to be identically
zero.  Consequently,
\begin{equation}\label{eq:final-rational-certificate}
  P_4
  =
  \sigma_0
  +\sum_i x_i\sigma_i
  +\sum_{i<j}x_ix_j\sigma_{ij}
  +hq.
\end{equation}
Remark that all SOS polynomials in \eqref{eq:final-rational-certificate} have rational
coefficients.  The positive definiteness of the recovered
matrices \(S_b^{\mathbb Q}\) is certified by exact
\(LDL^{\mathsf T}\) decompositions. Consequently, the reconstructed Gram
matrices are positive semidefinite.  An independent exact expansion verifies
\eqref{eq:final-rational-certificate} coefficientwise over \(\mathbb Q\).

The main SOS consists of \(152\) positively weighted rational squares, each
of the \(136\) small SOS blocks consists of \(16\) such squares, and \(q\)
is a cubic polynomial with rational coefficients.  Hence, 
\[
  P_4=P_{4,\bar\lambda}\in\mathcal T_2,
  \qquad
  \bar\lambda=\frac8{13},
\]
which proves the quantitative
stability result and hence \(DH_4\).  The exact certificate is formally
verified in Lean in Section~\ref{sec:lean-formalization}.

\section{Formal Verification in Lean}
\label{sec:lean-formalization}

Lean 4~\cite{Lean4} is an interactive theorem prover in which mathematical definitions, statements, and proofs are expressed in a formal language and checked mechanically. In this paper, rather than being used to discover the SOS certificate, Lean is used to verify the exact rational certificate constructed in Section~\ref{sec:agent-certificate} and derive machine-checked proofs of the two main theorems.  We first verify the quantitative stability estimate in Theorem~\ref{thm:stability}, from
which the Dittert conjecture in dimension \(4\) stated in
Theorem~\ref{thm:main} follows.

The formalization uses the mathematical library \texttt{Mathlib4}~\cite{Mathlib2020}, and the Lean environment is fixed by the toolchain
\texttt{v4.30.0-rc1}.  The source code is available at
\url{https://github.com/123ljh0bot/Dittert_Conjecture_in_Dimension_4}.

In Lean, a \(4\times4\) matrix is represented as a function taking a row
index and a column index in \(\mathrm{Fin}\,4\) and returning a real
number. The simplex constraint and the uniform matrix are encoded as
follows.

\begin{lstlisting}[language=lean]
abbrev Matrix4 := Fin 4 → Fin 4 → ℝ

def K4 : Set Matrix4 :=
  { A | (∀ i j, 0 ≤ A i j) ∧
    A 0 0 + A 0 1 + A 0 2 + A 0 3 +
    A 1 0 + A 1 1 + A 1 2 + A 1 3 +
    A 2 0 + A 2 1 + A 2 2 + A 2 3 +
    A 3 0 + A 3 1 + A 3 2 + A 3 3 = 4 }

def U4 : Matrix4 := fun _ _ => (1 / 4 : ℝ)
\end{lstlisting}

The Dittert functional \eqref{eq:phi} is encoded directly from the matrix
entries as follows.

\begin{lstlisting}[language=lean]
def φ_entries
    (a11 : ℝ) (a12 : ℝ) (a13 : ℝ) (a14 : ℝ)
    (a21 : ℝ) (a22 : ℝ) (a23 : ℝ) (a24 : ℝ)
    (a31 : ℝ) (a32 : ℝ) (a33 : ℝ) (a34 : ℝ)
    (a41 : ℝ) (a42 : ℝ) (a43 : ℝ) (a44 : ℝ) : ℝ :=
  row_product_entries a11 a12 a13 a14 a21 a22 a23 a24
    a31 a32 a33 a34 a41 a42 a43 a44
  + column_product_entries a11 a12 a13 a14 a21 a22 a23 a24
    a31 a32 a33 a34 a41 a42 a43 a44
  - permanent_entries a11 a12 a13 a14 a21 a22 a23 a24
    a31 a32 a33 a34 a41 a42 a43 a44
    
def φ (A : Matrix4) : ℝ :=
  φ_entries
    (A 0 0) (A 0 1) (A 0 2) (A 0 3)
    (A 1 0) (A 1 1) (A 1 2) (A 1 3)
    (A 2 0) (A 2 1) (A 2 2) (A 2 3)
    (A 3 0) (A 3 1) (A 3 2) (A 3 3)
\end{lstlisting}

Here the function \leanin{\ensuremath{\varphi}\_entries} evaluates the Dittert functional
directly from the sixteen matrix entries. The permanent is defined using
Mathlib's \leanin{Matrix.permanent}. A bridging lemma formally proves
that this entrywise definition agrees with \eqref{eq:phi}. The squared Frobenius
distance \eqref{eq:Delta4} is represented as
\leanin{\ensuremath{\|}A - U4\ensuremath{\|}\_F\^{}2}.

The SOS certificate is represented in Lean as a sparse polynomial with
rational coefficients. All polynomial operations  involved in the
verification  are performed exactly over \(\mathbb{Q}\), without
floating-point approximations. The gap polynomial \(Q_4\), the squared distance
\(\Delta_4\), and the simplex equality \(h\) are encoded from their
mathematical definitions and combined to form the target polynomial
\(13Q_4-8\Delta_4\) verified below.

\begin{lstlisting}[language=lean]
abbrev P := Poly 16

def dh4IntegerPoly : P :=
  polySum [const 61, scale (-32) rowProductPoly,
    scale (-32) columnProductPoly, scale 32 permanentPoly]

def d4Poly : P :=
  polySum [centeredSquare 0, centeredSquare 1, ... ,
    centeredSquare 15]

def simplexEquality : P :=
  sub (polySum [var 0, var 1, ... , var 15]) (const 4)

def compactTarget : P :=
  sub (scale 13 dh4IntegerPoly) (scale 8 d4Poly)
\end{lstlisting}
Here \leanin{dh4IntegerPoly}, \leanin{d4Poly}, and
\leanin{compactTarget} encode \(Q_4\), \(\Delta_4\), and
\(13Q_4-8\Delta_4=13P_4\), respectively. These identifications are
formally verified by the corresponding evaluation lemmas.

The certificate data consist of the main SOS block
\leanin{mainSquares}, the \(136\) small SOS blocks represented by
\leanin{templates} and \leanin{assignments}, and the cubic multiplier
\leanin{qPoly}.  These objects encode the rational SOS certificate
constructed in Section~\ref{sec:agent-certificate}.  
Their combination produces the reconstructed certificate polynomial
\leanin{rawCertificate} and the target polynomial
\leanin{compactTarget} verified below.

\begin{lstlisting}[language=lean]
def positiveCertificate : P :=
  add (sos mainSquares) (assignmentSum assignments)

def rawCertificate : P :=
  add positiveCertificate (mul simplexEquality qPoly)
\end{lstlisting}

The formal verification of the SOS certificate consists of two parts.
First, Lean verifies the nonnegativity of all rational weights appearing in
the SOS representation, ensuring that the reconstructed SOS terms
are nonnegative.

\begin{lstlisting}[language=lean]
theorem main_weights_nonnegative :
    WeightsNonnegative mainSquares := by
  apply weightsNonnegative_of_bool
  native_decide

theorem assignment_templates_weights_nonnegative :
    ∀ assignment ∈ assignments,
      WeightsNonnegative (templates.getD assignment.template []) := by
  ...
  native_decide
\end{lstlisting}

Second, Lean verifies the polynomial identity between the reconstructed
certificate and the target polynomial.  Since sparse polynomial
representations are not canonical, we compare their normalized forms.

\begin{lstlisting}[language=lean]
theorem normalized_certificate :
    FastPolyForSOS.fastNormalize rawCertificate = targetPoly := by
  native_decide

theorem normalized_compactTarget :
    FastPolyForSOS.fastNormalize compactTarget = targetPoly := by
  native_decide
\end{lstlisting}

The correctness theorem
\leanin{eval\_fastNormalize} shows that normalization preserves polynomial
evaluation. Therefore, the two normalization identities above imply that
\leanin{rawCertificate} and \leanin{compactTarget} have identical values at
every point of \(\mathbb{R}^{16}\).

It remains to derive the stability estimate from the verified certificate.
For \(A\in K_4\), the SOS terms in
\leanin{positiveCertificate} are nonnegative because all weights are
nonnegative and the generators \(x_i,x_ix_j\) are nonnegative on the
simplex. Since the simplex equality vanishes on \(K_4\), the multiplier
term vanishes, and therefore \leanin{rawCertificate} is nonnegative on
\(K_4\). By the verified polynomial identity,
\(\leanin{compactTarget}\) is nonnegative on \(K_4\). Therefore, the
verified certificate proves Theorem~\ref{thm:stability}.
Its formal statement is given below.

\begin{lstlisting}[language=lean]
theorem theorem_1_2_quantitative_stability_for_DH4
    (A : Matrix4) (hA : A ∈ K4) :
    (61 / 32 : ℝ) - φ A ≥ (1 / 52 : ℝ) * ‖A - U4‖_F^2
\end{lstlisting}

Consequently, Theorem~\ref{thm:main} follows from the quantitative stability estimate, whose formal statement is given below.

\begin{lstlisting}[language=lean]
theorem theorem_1_1_dittert_conjecture_dimension_four
    (A : Matrix4) (hA : A ∈ K4) :
    φ A ≤ 61 / 32 ∧ (φ A = 61 / 32 ↔ A = U4)
\end{lstlisting}

\section*{Acknowledgments}

The authors thank Professor Lihong Zhi for helpful comments and suggestions. This work was supported in part by the National Key Research and Development Program of China under Grant 2023YFA1009402.

\bibliographystyle{elsarticle-num}
\bibliography{references}

\end{document}

%% file: lstlean.tex
\providecolor{keywordcolor}  {HTML}{4B69C6}  
\providecolor{tacticcolor}   {HTML}{4B69C6}  
\providecolor{declcolor}     {HTML}{AA3731}  
\providecolor{identcolor}    {HTML}{333333}  
\providecolor{sortcolor}     {HTML}{333333}  
\providecolor{symbolcolor}   {HTML}{333333}  
\providecolor{stringcolor}   {HTML}{448C27}  
\providecolor{attributecolor}{HTML}{7A3E9D}  
\providecolor{commentcolor}  {HTML}{8A8A8A}  
\providecolor{numbercolor}   {HTML}{9C5D27}  
\providecolor{springcolor}   {HTML}{333333}  
\providecommand{\uplambda}{\lambda}
\providecommand{\Uppi}{\Pi}

\lstdefinelanguage{lean} {
mathescape=false,
texcl=false,
morekeywords=[1]{
import, prelude, protected, private, noncomputable, definition, meta, renaming,
hiding, parameter, parameters, begin, constant, constants,
lemma, variable, variables, theory,
print, theorem, example,
open, as, export, override, axiom, axioms, inductive, with,
structure, record, universe, universes,
alias, help, precedence, reserve, declare_trace, add_key_equivalence,
match, infix, infixl, infixr, notation, postfix, prefix, instance,
eval, reduce, check, end, this,
using, using_well_founded, namespace, section,
attribute, local, set_option, extends, include, omit, class,
raw, replacing,
calc, have, show, suffices, by, in, at, let, forall, Pi, fun,
exists, if, dif, then, else, assume, obtain, from, register_simp_ext, unless, break, continue,
mutual, do, def, run_cmd, const,
partial, mut, where, macro, syntax, deriving,
return, try, catch, for, macro_rules, declare_syntax_cat, abbrev},
morekeywords=[2]{Sort, Type, Prop},
morekeywords=[5]{Matrix4, K4, U4, phi_entries, row_product_entries,
column_product_entries, permanent_entries, frobenius_norm_squared,
theorem_1_1_dittert_conjecture_dimension_four,
theorem_1_2_quantitative_stability_for_DH4,
dh4IntegerPoly, d4Poly, simplexEquality, compactTarget,
positiveCertificate, rawCertificate, normalized_certificate,
normalized_compactTarget, main_weights_nonnegative},
morekeywords=[3]{
assumption,
apply, intro, intros, allGoals,
generalize, clear, revert, done, exact,
refine, repeat, cases, rewrite, rw,
simp, simp_all, contradiction,
constructor, injection,
induction,
},
literate=
{α}{{\ensuremath{\mathrm{\alpha}}}}1
{β}{{\ensuremath{\mathrm{\beta}}}}1
{γ}{{\ensuremath{\mathrm{\gamma}}}}1
{δ}{{\ensuremath{\mathrm{\delta}}}}1
{ε}{{\ensuremath{\mathrm{\varepsilon}}}}1
{ζ}{{\ensuremath{\mathrm{\zeta}}}}1
{η}{{\ensuremath{\mathrm{\eta}}}}1
{θ}{{\ensuremath{\mathrm{\theta}}}}1
{ι}{{\ensuremath{\mathrm{\iota}}}}1
{κ}{{\ensuremath{\mathrm{\kappa}}}}1
{μ}{{\ensuremath{\mathrm{\mu}}}}1
{ν}{{\ensuremath{\mathrm{\nu}}}}1
{ξ}{{\ensuremath{\mathrm{\xi}}}}1
{π}{{\ensuremath{\mathrm{\mathnormal{\pi}}}}}1
{ρ}{{\ensuremath{\mathrm{\rho}}}}1
{σ}{{\ensuremath{\mathrm{\sigma}}}}1
{τ}{{\ensuremath{\mathrm{\tau}}}}1
{φ}{{\ensuremath{\mathrm{\varphi}}}}1
{χ}{{\ensuremath{\mathrm{\chi}}}}1
{ψ}{{\ensuremath{\mathrm{\psi}}}}1
{ω}{{\ensuremath{\mathrm{\omega}}}}1
{Γ}{{\ensuremath{\mathrm{\Gamma}}}}1
{Δ}{{\ensuremath{\mathrm{\Delta}}}}1
{Θ}{{\ensuremath{\mathrm{\Theta}}}}1
{Λ}{{\ensuremath{\mathrm{\Lambda}}}}1
{Φ}{{\ensuremath{\mathrm{\Phi}}}}1
{Ξ}{{\ensuremath{\mathrm{\Xi}}}}1
{Ψ}{{\ensuremath{\mathrm{\Psi}}}}1
{Ω}{{\ensuremath{\mathrm{\Omega}}}}1
{ℵ}{{\ensuremath{\aleph}}}1
{≤}{{\ensuremath{\leq}}}1
{≥}{{\ensuremath{\geq}}}1
{≠}{{\ensuremath{\neq}}}1
{≈}{{\ensuremath{\approx}}}1
{≡}{{\ensuremath{\equiv}}}1
{≃}{{\ensuremath{\simeq}}}1
{∂}{{\ensuremath{\partial}}}1
{∆}{{\ensuremath{\triangle}}}1 
{∫}{{\ensuremath{\int}}}1
{∑}{{\ensuremath{\mathrm{\Sigma}}}}1
{⊥}{{\ensuremath{\perp}}}1
{∞}{{\ensuremath{\infty}}}1
{∓}{{\ensuremath{\mp}}}1
{±}{{\ensuremath{\pm}}}1
{×}{{\ensuremath{\times}}}1
{⊕}{{\ensuremath{\oplus}}}1
{⊗}{{\ensuremath{\otimes}}}1
{⊞}{{\ensuremath{\boxplus}}}1
{∇}{{\ensuremath{\nabla}}}1
{√}{{\ensuremath{\sqrt}}}1
{⬝}{{\ensuremath{\cdot}}}1
{•}{{\ensuremath{\cdot}}}1
{∘}{{\ensuremath{\circ}}}1
{⁻}{{\ensuremath{^{-}}}}1
{∧}{{\ensuremath{\wedge}}}1
{∨}{{\ensuremath{\vee}}}1
{¬}{{\ensuremath{\neg}}}1
{⊢}{{\ensuremath{\vdash}}}1
{⟨}{{\ensuremath{\langle}}}1
{⟩}{{\ensuremath{\rangle}}}1
{↦}{{\ensuremath{\mapsto}}}1
{←}{{\ensuremath{\leftarrow}}}1
{<-}{{\ensuremath{\leftarrow}}}1
{→}{{\ensuremath{\rightarrow}}}1
{↔}{{\ensuremath{\leftrightarrow}}}1
{⇒}{{\ensuremath{\Rightarrow}}}1
{⟹}{{\ensuremath{\Longrightarrow}}}1
{⇐}{{\ensuremath{\Leftarrow}}}1
{⟸}{{\ensuremath{\Longleftarrow}}}1
{∩}{{\ensuremath{\cap}}}1
{∪}{{\ensuremath{\cup}}}1
{⊂}{{\ensuremath{\subseteq}}}1
{⊆}{{\ensuremath{\subseteq}}}1
{⊄}{{\ensuremath{\nsubseteq}}}1
{⊈}{{\ensuremath{\nsubseteq}}}1
{⊃}{{\ensuremath{\supseteq}}}1
{⊇}{{\ensuremath{\supseteq}}}1
{⊅}{{\ensuremath{\nsupseteq}}}1
{⊉}{{\ensuremath{\nsupseteq}}}1
{∈}{{\ensuremath{\in}}}1
{∉}{{\ensuremath{\notin}}}1
{∋}{{\ensuremath{\ni}}}1
{∌}{{\ensuremath{\notni}}}1
{∅}{{\ensuremath{\emptyset}}}1
{∖}{{\ensuremath{\setminus}}}1
{†}{{\ensuremath{\dag}}}1
{ℕ}{{\ensuremath{\mathbb{N}}}}1
{ℤ}{{\ensuremath{\mathbb{Z}}}}1
{ℝ}{{\ensuremath{\mathbb{R}}}}1
{ℚ}{{\ensuremath{\mathbb{Q}}}}1
{ℂ}{{\ensuremath{\mathbb{C}}}}1
{⌞}{{\ensuremath{\llcorner}}}1
{⌟}{{\ensuremath{\lrcorner}}}1
{⦃}{{\ensuremath{\{\!|}}}1
{⦄}{{\ensuremath{|\!\}}}}1
{‖}{{\ensuremath{\|}}}1
{₁}{{\ensuremath{_1}}}1
{₂}{{\ensuremath{_2}}}1
{₃}{{\ensuremath{_3}}}1
{₄}{{\ensuremath{_4}}}1
{₅}{{\ensuremath{_5}}}1
{₆}{{\ensuremath{_6}}}1
{₇}{{\ensuremath{_7}}}1
{₈}{{\ensuremath{_8}}}1
{₉}{{\ensuremath{_9}}}1
{₀}{{\ensuremath{_0}}}1
{ᵢ}{{\ensuremath{_i}}}1
{ⱼ}{{\ensuremath{_j}}}1
{ₐ}{{\ensuremath{_a}}}1
{¹}{{\ensuremath{^1}}}1
{ₙ}{{\ensuremath{_n}}}1
{ₘ}{{\ensuremath{_m}}}1
{ₚ}{{\ensuremath{_p}}}1
{↑}{{\ensuremath{\uparrow}}}1
{↓}{{\ensuremath{\downarrow}}}1
{...}{{\ensuremath{\ldots}}}1
{·}{{\ensuremath{\cdot}}}1
{▸}{{\ensuremath{\triangleright}}}1
{Σ}{{\color{symbolcolor}\ensuremath{\Sigma}}}1
{Π}{{\color{symbolcolor}\ensuremath{\Uppi}}}1 
{∀}{{\color{symbolcolor}\ensuremath{\forall}}}1
{∃}{{\color{symbolcolor}\ensuremath{\exists}}}1
{λ}{{\color{symbolcolor}\ensuremath{\uplambda}}}1
{\$}{{\color{symbolcolor}\$}}1
{:=}{{\color{identcolor}:=}}1
{=}{{\color{identcolor}=}}1
{<|>}{{\color{symbolcolor}<|>}}1
{<\$>}{{\color{symbolcolor}<\$>}}1
{+}{{\color{identcolor}+}}1
{*}{{\color{identcolor}*}}1,
morecomment=[s][\color{commentcolor}]{/-}{-/},
morecomment=[l][\itshape \color{commentcolor}]{--},
showstringspaces=false,
keepspaces=true,
morestring=[b]",
morestring=[d],
tabsize=3,
extendedchars=true,
sensitive=true,
breaklines=true,
breakatwhitespace=true,
basicstyle=\ttfamily\small,
captionpos=b,
columns=[l]fullflexible,
frame = shadowbox,
escapeinside=``,
xleftmargin=2em,
xrightmargin=2em, 
aboveskip=1em,
identifierstyle={\ttfamily\color{identcolor}},
keywordstyle=[1]{\ttfamily\color{keywordcolor}},
keywordstyle=[2]{\ttfamily\color{sortcolor}},
keywordstyle=[3]{\ttfamily\color{tacticcolor}},
keywordstyle=[4]{\ttfamily\color{attributecolor}},
keywordstyle=[5]{\ttfamily\bfseries\color{declcolor}},
stringstyle={\ttfamily\color{stringcolor}},
commentstyle={\ttfamily\footnotesize\color{commentcolor}},
    numbers=left, 
    numberstyle=\tiny\color{gray}, 
    stepnumber=1, 
    numbersep=6pt, 
    firstnumber=1, 
}